\documentclass[jap,graphicx, reprint,unsortedaddress]{revtex4-1} 
\usepackage{amsfonts, amsmath, graphicx, feynmf}
\usepackage[version=3]{mhchem}

\begin{document}

\title{Effective Hamiltonian for surface states of topological insulator nanotubes} 
\author{Zhuo Bin Siu}
\affiliation{Computational Nanoelectronics and Nanodevices Laboratory, Electrical and Computer Engineering Department, National University of Singapore, Singapore} 
\author{Seng Ghee Tan} 
\affiliation{Data Storage Institute, Agency for Science, Technology and Research (A*STAR), Singapore} 
\author{Mansoor B. A. Jalil} 
\affiliation{Computational Nanoelectronics and Nanodevices Laboratory, Electrical and Computer Engineering Department, National University of Singapore, Singapore} 

\begin{abstract}
In this work we derive an effective Hamiltonian for the surface states of a hollow topological insulator (TI) nanotube with finite width walls. Unlike a solid TI cylinder, a TI nanotube possesses both an inner as well as outer surface on which the states localized at each surface are coupled together. The curvature along the circumference of the nanotube leads to a spatial variation of the spin orbit interaction field experienced by the charge carriers as well as an asymmetry between the inner and outer surfaces of the nanotube. Both of these features result in terms in the effective Hamiltonian for a TI nanotube absent in that of a \textit{flat} TI thin film of the same thickness. We calculate the numerical values of the parameters for a \ce{Bi2Se3} nanotube as a function of the inner and outer radius, and show that the differing relative magnitudes between the parameters result in qualitatively differing behaviour for the eigenstates of tubes of different dimensions.  	
\end{abstract} 

\maketitle

\section{Introduction} 

\begin{figure}[ht!]
\centering
\includegraphics[scale=0.35]{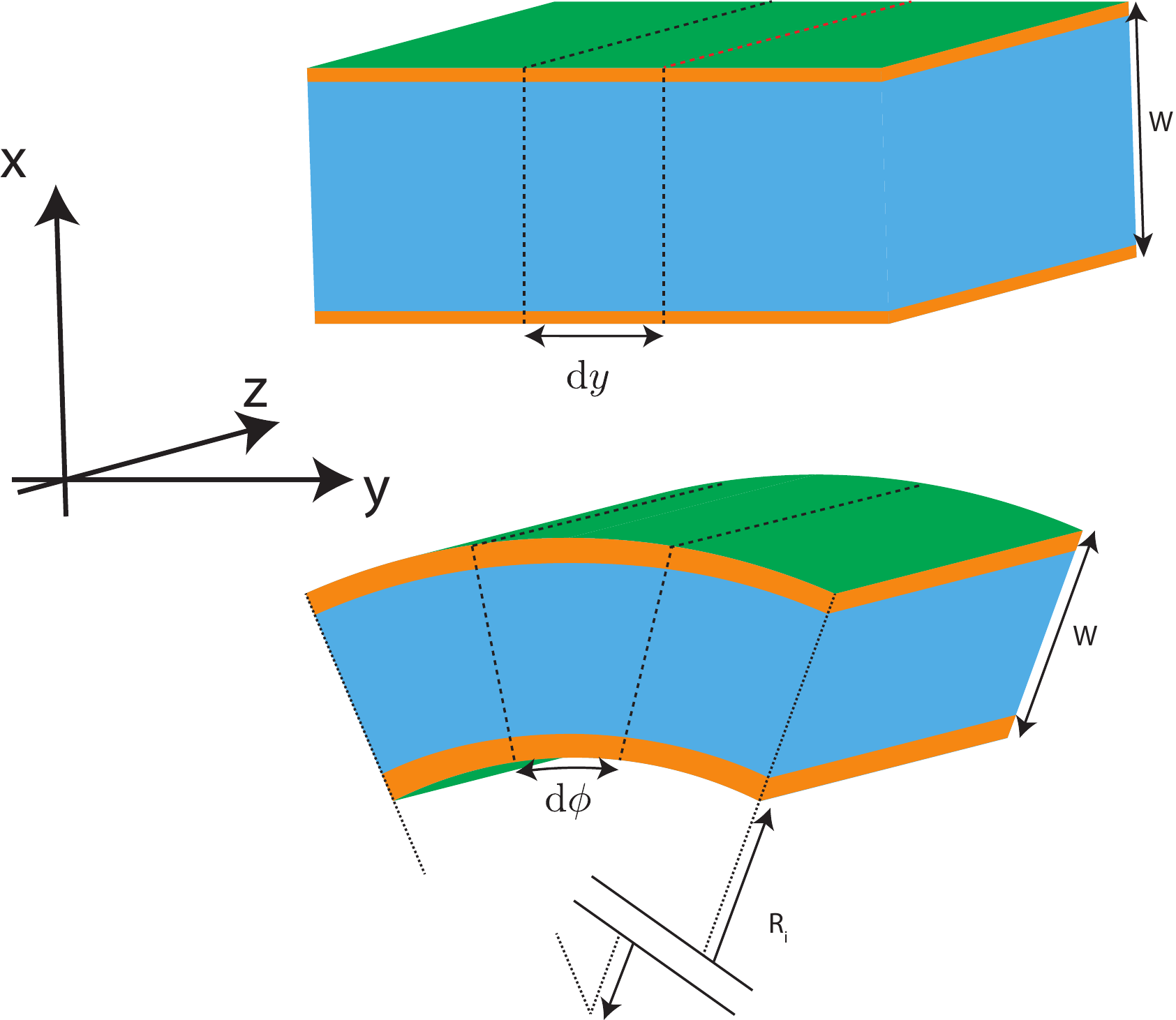}
\caption{  A section of a flat TI thin film  of width $W$ (top)  and a TI nanotube of thickness $W$ and inner radius $R_i$ (bottom). The flat thin film on the top extends to infinity along the $y$ and $z$ directions and has finite thickness along the $x$ direction.  The nanotube section extends to infinity in the $z$ direction and has finite thickness along the radial $r$ direction. The orange colored segments represent schematically the regions where the surface states considered in this paper are localized around.  The infinitesimal cross section elements illustrate that whereas the flat thin film is isotropic across the thickness for a $\mathrm{d}$y slice,  there is an asymmetry between the inner and outer radius of the nanotube for a $\mathrm{d}\phi$ slice.   } 

\label{gSchemA}
\end{figure}

Topological insulators (TI) are an emerging class of materials which have attracted much attention due to the unique properties of their surface states \cite{RMP82_3045}. In particular, topological insulator thin films have been studied by various authors \cite{PRB80_205401,PRB81_041307,PRB81_115407}. 
A key feature that distinguishes a TI thin film (Fig. \ref{gSchemA} top)  from a TI slab of semi-infinite thickness is that there are now two surfaces, which we label as the top and bottom surfaces, each of which admits states localized at the respective surfaces. The finite thickness of the films leads to a coupling between the top and bottom surfaces.  The states localized on the top and bottom surfaces are not independent of each other but are correlated, for example by the boundary condition that their wavefunctions have to simultaneously vanish at both surfaces. 

We now ask the question of what happens when we introduce curvature into the system by considering the specific example of a TI nanotube (Fig. \ref{gSchemA} bottom)  with walls of finite uniform thickness and its axis parallel to the quintuple layer normal. The study of curvature in TI thin films is motivated by the fact that strong spin-orbit coupling in TI systems enable the control of either the spin or momentum degree of freedom to control the other. One way of manipulating the momentum direction of the charge carriers is to confine them to move on curved surfaces so that the momentum direction of the charge carriers change as they move along the surfaces. The manipulation of spin by curvature in curved TI systems gives rise to interesting effects that may be of technological application. For example, we showed in an earlier paper \cite{JAP117_17C749} that in \textit{solid} TI cylinders (which have only an outer surface),  an anomalous magnetoresistance behavior emerges in the transmission between two TI cylinders magnetized in different directions perpendicular to the cylinder axis because of the position dependence of the spin orbit interaction field around the circumference of the cylinder. 

In a TI nanotube the presence of both an inner and outer surface results in an additional degree of freedom which can be related to which of the two surfaces a surface state is localized at. This additional degree of freedom yields richer physics for the TI nanotube system compared to the TI cylinder system.      

Compared to the flat film (Fig. \ref{gSchemA} top), the nanotube has two main differences.  First, the breaking of the symmetry between the inner and outer radius of the cylinder leads to the emergence of terms in the Hamiltonian which cancel out and vanish on the flat slab.  Second, the SOI field on a TI surface lies tangential to the surface. The presence of curvature leads to a position dependence of the SOI field on the angular position along the circumference of the tube. Both of these manifest as the emergence of more terms in the effective surface state Hamiltonian for the surface states of a TI nanotube, whose derivation will be the main focus of this paper. 

In this paper we derive the effective Hamiltonian for the surface states of a TI nanotube. We also derive, in parallel, the corresponding effective surface state Hamiltonian for a flat TI thin film. The comparison between the two illustrates the effects of curvature in a TI thin film. To further elucidate the properties of the cylinder surface states we next calculate the lowest energy eigenstates for some values of the nanotube wall thickness $W$ and inner radius $R_i$ using the derived effective Hamiltonian.

\section{ The Liu 4-band Hamiltonian} 
Our starting point is the effective four-band model Hamiltonian of Liu \textit{et al} \cite{PRB82_045122} which describes both the bulk and surface states of the BiSe family of topological insulators near the $k$-space $\Gamma$ point. The Hamiltonian reads 
\begin{eqnarray}
	H_{(4B)} &=& (C_0 + C_1 k_z^2 + C_2 k_\parallel^2)  \nonumber \\
	&& + (M_0 + M_1 k_z^2 + M_2 k_\parallel^2)\Gamma_5 + B\Gamma_4k_z   \nonumber \\ 
	&&  + A(\Gamma_1k_y-\Gamma_2k_x) \label{HCa} 
\end{eqnarray}
where
\begin{eqnarray*}
	&& \Gamma_1 = \sigma_x\otimes t_1,	\Gamma_2 = \sigma_y\otimes t_1, \\
	&& \Gamma_3 = \sigma_z\otimes t_1, \Gamma_4 = \mathbb{I}_\sigma\otimes t_2, 	\Gamma_5 = \mathbb{I}_\sigma\otimes t_3, 
\end{eqnarray*}
and $k_\parallel^2 \equiv k_x^2+k_y^2$. The $t$'s can be interpreted as describing an orbital degree of freedom and the $\sigma$'s the real spins. Our approach for both the flat film as well as nanotube follows that of Lu \textit{et al.} \cite{PRB81_115407}.   We separate Eq. \ref{HCa} into two parts -- a `perpendicular Hamiltonian' $H_{(4B),\perp}$ containing constant terms and derivatives perpendicular to the two surfaces, and the remaining `parallel Hamiltonian' $H_{(4B),\parallel}$ containing derivatives tangential to the two surfaces. We first solve for the energy eigenstates of the perpendicular Hamiltonian that decay exponentially away from the two surfaces. These states are hence localized around the surfaces and represent the surface states which we seek. The effective Hamiltonian for the surface states are then obtained by treating $H_{(4B),\parallel}$ as a perturbation to $H_{(4B),\perp}$. The localized eigenstates of $H_{(4B),\perp}$ are at least two-fold degenerate due to spin degeneracy.  Consistent with standard degenerate perturbation theory, treating $H_{(4B),\parallel}$ as a perturbation amounts to projecting $H_{(4B),\parallel}$ in the basis of the degenerate eigenstates of $H_{(4B), \perp}$.

\section{Perpendicular Hamiltonian}
\subsection{Flat TI  thin film}  

We first consider the flat TI thin film, for which some analytic expressions can be obtained.  We shall later see that the localized perpendicular Hamiltonian eigenstates of a TI nanotube can, to a very good approximation, be related to those of the flat film.  To make a fair comparison with the TI cylinder with axis along the $z$ direction, we consider a flat TI film with its normal along the $x$ direction so that in both of these systems, we have one direction on the surface parallel to the TI quintuple layer plane and an orthogonal direction on the surface perpendicular to the quintuple plane. Note that this differs from the usual flat TI thin films considered in earlier works where both in-plane directions are parallel to the quintuple plane.  
  
 $H_{(4B),\perp}$ in the flat thin film containing constant terms and the $x$ derivatives reads 
\begin{equation}
	C_0 +C_2k_x^2 + A_0 k_x  t_x \sigma_y  + M t_z + M_2 t_z k_x^2.
\end{equation}
The real spin degree of freedom is diagonalized by the eigenstates of $\sigma_y$ which we denote as $|\pm \sigma_y\rangle$ 	  

For the $|\pm_{(\sigma_y)} \sigma_y\rangle$ states ( the ${(\sigma_y)}$ subscript indicates that the $\pm$ pertains to the $\sigma_y$ degree of freedom in order to distinguish this from the other $\pm$s which will occur later),  we have 
\begin{equation} 
	C_0 +C_2k_x^2 \pm_{(\sigma_y)} A_0 k_x  t_x  + M t_z + M_2 t_z k_x^2. \label{E0} 
\end{equation} 

Since we are looking for localized states, we search for states with the form of $\exp(\lambda x)$ , so that $k_x \rightarrow -i \lambda$.  For a given eigenenergy $E_f$, diagonalizing Eq. \ref{E0} and equating the eigenenergies with $E_f$ give an quadratic equation in $\lambda^2$. Denoting the two solutions of the quadratic equation as $\lambda_{\pm_{(\lambda)}}^2$, we have
 
\begin{eqnarray}
&& \lambda_{\pm_{(\lambda)}}^2 = \nonumber \\
&& -\frac{1}{ 2(C_2^2 - M_2^2)} \Big(A_0^2 - 2C_0C_2 + 2C_2E_f + 2MM_2 \pm_{(\lambda)} \nonumber \\
&& \big(A_0^4 + 4(C_2M - C_0M_2 + E_fM_2)^2 \nonumber \\
&& + A_0^2(-4C_0C_2 + 4C_2E_f + 4MM_2) \big)^{\frac{1}{2}}\Big) \label{L2Eq} 
\end{eqnarray}

We seek linear combinations of these exponentials which disappear simultaneously at the two surfaces at $x=\pm W/2$. Two such linearly independent combinations are 
\begin{eqnarray*}
	f_+ &\equiv&  \frac{\cosh(\lambda_+ x)}{\cosh(\lambda_+ W / 2)} - \frac{\cosh(\lambda_- x)}{\cosh(\lambda_- W / 2)}\\
	f_- &\equiv&  \frac{\sinh(\lambda_+ x)}{\sinh(\lambda_+ W/ 2)} - \frac{\sinh(\lambda_- x)}{\sinh(\lambda_- W / 2)}.
\end{eqnarray*}

The $f_+$ has even parity whereas $f_-$ has odd parity. In order to diagonalize Eq. \ref{E0} for each of the two values of $\pm_{(\sigma_y)}=+1$ or $-1$, we only need to consider  (in the usual Pauli matrix representation of $t_x = \begin{pmatrix} 0 & 1 \\ 1 & 0 \end{pmatrix}$) the following combinations
\[
	\chi_{\pm_{\sigma_y}}  \equiv \begin{pmatrix} f_+ \\ c_{-, \pm_{\sigma_y}} f_- \end{pmatrix} , \varphi_{\pm_{\sigma_y}}  \equiv \begin{pmatrix} f_- \\ c_{+,\pm_{\sigma_y}} f_+  \end{pmatrix}
\] 

Substituting, for example, $\chi_{\pm_{\sigma_y}}$ into $(H-E_f)\chi_{\pm_{\sigma_y}} = 0$ gives a set of 2 equations which contain hyperbolic trigonometric functions of $x$  but which should nonetheless give 0 everywhere independent of the value of $x$. This indicates the coefficients in front of the various hyperbolic trigonometric functions should go to 0. Thus, setting the coefficient of $\cosh(\lambda_+ x)$ in the upper component of $(H-E_f)\chi_{\pm_{\sigma_y}} = 0$ to 0 gives one expression for $c_{-,\pm_{\sigma_y}}$  while setting the coefficient of $\cosh(\lambda_- x)$ to 0 gives another expression for $c_{-,\pm_{\sigma_y}}$. Imposing the condition that these two expressions  for $c_{-,\pm_{\sigma_y}}$ agree yields the equation 
\begin{equation}
	\frac{ ( C_0 - E_f + M - (C_2 + M_2) \lambda_+^2) \lambda_-\tanh(W \lambda_+ / 2)}{ (C_0 - E_f + M - (C_2 + M_2)\lambda_-^2)\lambda_+ \tanh(W \lambda_-/2)} = 1. \label{phiDetEq} 
\end{equation}

This is essentially a transcendental equation in $E_f$ due to the $E_f$ dependence of  $\lambda_{\pm_{(\lambda)}}$ via Eq. \ref{L2Eq}. The equation can be solved numerically.  Eq. \ref{E0} from which the equation is derived differs only in the sign of the $A_0$ term for the two possible values of $\pm_{(\lambda)}$. $A_0$ however does not appear explicitly in Eq. \ref{phiDetEq} above and only appears in even powers in the $\lambda_{\pm_{(\lambda)}}$s. The $\chi_{\pm_{\sigma_y}}$ states are thus degenerate. We denote the energy of these states as $E_\chi$.

Once we find an energy where the values of $c_{-, \pm{\sigma_y}}$ calculated from the equations resulting coefficients of $\cosh(\lambda_{\pm_{(\lambda)}} x)$ agree, we can use either expression to obtain the value of $c_{-,\pm{\sigma_y}}$. A similar procedure can be applied on  $(H-E_f)\varphi_{\pm_{\sigma_y}} = 0$ to obtain the corresponding eigenenergy $E_\varphi$ and eigenspinor. 

\subsection{Cylindrical nanotube} 
We now proceed to derive the perpendicular Hamiltonian for the TI nanotube. We rewrite Eq. \ref{HCa} in cylindrical coordinates using $k_x^2+k_y^2 = - ( \partial_r^2 + \frac{1}{r}\partial_r + \frac{1}{r^2}\partial_\phi^2) \equiv \nabla^2_{r\phi}$ (we included the $r\phi$ subscript in the $\nabla^2$ to distinguish it from the full Laplacian operator which has an additional $\partial_z^2$ term), as well as $k_x = -i\partial_x = -i ( \frac{\partial r}{\partial x}\partial_r + \frac{\partial \phi}{\partial x}\partial_\phi)$ and its analog for $k_y$. Denoting this as $H_{(4B),cy}$ with $cy$ for \textit{cy}lindrical, we have at $k_z=0$, 
\begin{eqnarray} 
	 && H_{(4B), cy}(k_z = 0) \nonumber\\
		&=& \mathbb{I}_4(C_0 - C_2 \nabla^2_{r\phi}) + M_0 \mathbb{I}_\sigma t_z \nonumber \\
		&&- M_2 \mathbb{I}_\sigma t_z \nabla^2_{r\phi} + A ( \sigma_r t_x \frac{k_\phi}{r} - \sigma_\phi t_x k_r)
\label{H4Bcy0} 
\end{eqnarray} 
where $k_r \equiv -i\partial_r$ and $k_\phi \equiv -i\partial_\phi$. 

This has a $A \sigma_\phi t_x k_r$ term which goes into our expression for $H_{4B,\perp,cy}$ but is inconvenient because $\sigma_\phi \equiv -\sin(\phi)\sigma_y+\cos(\phi)\sigma_x$ is dependent on the $\phi$ coordinate.  For later convenience, we therefore first diagonalize the spin degree of freedom by performing the unitary transformation 
\begin{equation}
	U = \frac{1}{\sqrt{2}} \begin{pmatrix} 1 & 1 \\ i\exp(i\phi) & -i\exp(i \phi) \end{pmatrix} \label{uTrans}
\end{equation}
so that
\[
	U^\dagger \sigma_r U =  \tilde{\sigma}_y, U^\dagger \sigma_\phi U =\tilde{\sigma}_z, U^\dagger \sigma_z U = \tilde{\sigma}_x. 
\]

Mathematically, the unitary transformation corresponds to a rotation of the spin axes so that the $\tilde{\sigma}_x$ now points along the $\sigma_\phi$ direction. For convenience we call the $\tilde{\vec{\sigma}}$ the `rotated frame', and the frame before the rotation the `lab frame'. The tilde on the operators on the right hand side reminds us that while the numerical representation of the operators  are the same 2 by 2 numerical matrices as the usual Pauli matrices, they are to be understood to be operators in the rotated frame. $U$ does not commute with $k_\phi \equiv -i\partial_\phi$ so that on performing $UH_{(4B),cy}U^\dagger$ we have additional terms emerging from the $k_\phi$ terms. We have, for the term in $H_{(4B), cy}$ containing $k_\phi$ and $k_r$, 
\begin{eqnarray} 
	&& U^\dagger A ( \sigma_r t_x \frac{k_\phi}{r} -\sigma_\phi t_x k_r) U|\Psi\rangle \nonumber\\
	&=& \frac{A}{r} t_x (\tilde{\sigma}_y k_\phi +  U^\dagger \sigma_r(-i\partial_\phi U) - r \tilde{\sigma}_z k_r)|\Psi\rangle \nonumber \\
	&=&\frac{A}{r} t_x(\tilde{\sigma}_y k_\phi  -(r k_r - \frac{i}{2})\tilde{\sigma}_z + \frac{1}{2}\tilde{\sigma}_y)|\Psi\rangle. \label{UkrU}
\end{eqnarray}

The emergence of the imaginary $-\frac{i}{2}\sigma_z$ term may seem alarming. This term is, however, a necessary ingredient in ensuring that the perpendicular Hamiltonian in cylindrical coordinates is Hermitian. The standard criteria for an arbitrary operator $O$ being Hermitian is that for $|\Psi\rangle$ and $|\Phi\rangle$ being arbitrary states, $\langle \Psi|O|\Phi\rangle = \langle \Phi|O|\Psi\rangle^*$. In cylindrical coordinates, this becomes $\int \mathrm{d}r \mathrm{d}\phi\ r \Psi^*O\Phi = \int \mathrm{d}r \mathrm{d}\phi\ r \Phi^*O\Psi$ in which there is an additional factor of $r$ in the integrand. According to this criteria, $-i\partial_r$ by itself is not Hermitian, but $-i(\partial_r + \frac{1}{2r})$ is. (The additional  $\frac{1}{2r}$ is in fact $\partial_r \ln\sqrt{g}$, $g$ being the determinant of the metric tensor. )   A physical $H_{(4B),\perp,cy}$ hence has to contain $-i(\partial_r + \frac{1}{2r})$ rather than $-i\partial_r$. The $-\frac{i}{2}\sigma_z$ term that appears thus gives the desired combination of $-i (\partial_r + \frac{1}{2r})$ required for Hermitricity. The unitary transformation also gives an additional factor of $\frac{A}{2r}t_x \tilde{\sigma}_y$ which we will exclude from the perpendicular Hamiltonian, and account for later in the parallel Hamiltonian. 

Performing the unitary transformation on Eq. \ref{H4Bcy0} gives a block diagonal matrix  with the upper diagonal block acting on the (lab frame) spin $+\sigma_\phi$ states,  given by 
\begin{eqnarray}
H_{\perp,+\sigma_\phi} &=& C_0 \mathbb{I}_t  + M_0 t_z \nonumber \\
&&  - (\partial^2_r+\frac{1}{r}\partial_r)(C_2 \mathbb{I}_t + t_z) + At_x i(\partial_r + \frac{1}{r}) 	\label{Hperpsig} 
\end{eqnarray}
and a lower diagonal block  $H_{\perp,-\sigma_\phi}$ acting the spin $-\sigma_\phi$ states. The lower block is related to the upper block via $H_{\perp,-\sigma_\phi} = U'H_{\perp,+\sigma_\phi} U'^\dagger$ with $U'=U'^\dagger=\sigma_z$.  $U'$ introduces a net $\pi$ phase difference between the $\pm t$ components of the eigenstate. This is in direct analog to the $|\varphi_{-\sigma_y}\rangle$ and  $|\chi_{-\sigma_y}\rangle$ states for the \textit{flat} thin film differing from  $|\varphi_{+\sigma_y}\rangle$ and  $|\chi_{+\sigma_y}\rangle$ respectively by having a net phase difference of $\pi$ between the $\pm t$ components. 

Eq. \ref{Hperpsig} does not admit a simple analytic solution. We thus find the eigenstates of Eq. \ref{Hperpsig} numerically, and employ the unitary transform $U'$ to obtain the eigenstate of $H_{\perp, -\sigma_\phi}$ from the eigenstate of $H_{\perp,+\sigma_\phi}$. For all the numerical results which follow, we use the material parameters for \ce{Bi2Se3} from Ref. \onlinecite{PRB82_045122} . 

\subsection{Relationship between flat film and nanotube perpendicular Hamiltonian eigenstates} 
The eigenstates or $H_{\perp,+\sigma_\phi}$ in the large $r$ limit are approximately related to those of the perpendicular Hamiltonian for a flat thin film, Eq. \ref{E0}, in the following sense. Let us denote the wavefunction of an eigenstate of Eq. \ref{Hperpsig} as $\psi$ so that $H_{\perp,+\sigma_\phi} \psi = E \psi$.  Dropping the terms in $H_{\perp,+\sigma_\phi}$ in Eq. \ref{Hperpsig} proportional to $\frac{1}{r}$, we have   

\begin{eqnarray*}
H_{\perp,+\sigma_\phi} (\sqrt{r}\psi) &\approx& \big(  C_0 \mathbb{I}_t  + M_0 t_z \nonumber \\
&&  - (\partial^2_r)(C_2 \mathbb{I}_t + t_z) + At_x i(\partial_r ) \big) (\sqrt{r}\psi).
\end{eqnarray*}
This  corresponds to $H_{\perp}$ for a flat TI thin film, Eq. \ref{E0},  with the identification of $\partial_r \rightarrow \partial_x$. We also have, dropping terms with inverse powers of $r$ larger than $1/2$, 
\begin{eqnarray*}
	&& H_{\perp,+\sigma_\phi}(\sqrt{r}\psi) \\
	&=& \sqrt{r} H_{\perp,+\sigma_\phi}\psi - (C_2 \mathbb{I}_t + M_2 t_z) (\frac{1}{4r^{3/2}}\psi \\
		&&+ \frac{1}{\sqrt{r}}\psi') + \frac{3iA}{2\sqrt{r}}\sigma_r\psi \\
		&\approx& \sqrt{r}H_{\perp,+\sigma_\phi}\psi \\
		&=&E (\sqrt{r}\psi). 
\end{eqnarray*} 

The eigenstates of the \textit{cylindrical} perpendicular Hamiltonian multiplied by $\sqrt{r}$, are thus approximately the eigenstates of the \textit{flat} perpendicular Hamiltonian of the same thickness and have approximately the same eigenenergy. These approximations are ultimately justified by a comparison between the exact wavefunctions obtained by solving Eq. \ref{Hperpsig} explicitly multiplied by $\sqrt{r}$, and the wavefunctions for a flat thin film of the same width.  A visual inspection (not shown) indicates that the wavefunctions cannot be distinguished apart by eye, even for the smallest value of $R_i = 5\ \mathrm{nm}$ and cylinder wall width $W= 100\ \mathrm{nm}$  considered in this paper.  We hence borrow the notation of $|\varphi_{\pm \sigma_\phi} \rangle$ and $|\chi_{\pm \sigma_\phi}\rangle $ to denote the eigenstates of the cylindrical perpendicular Hamiltonian to denote the states whose wavefunctions multiplied by $\sqrt{r}$ resemble those of the flat thin film $|\varphi_{\pm \sigma_y}\rangle$ and $|\chi_{\pm \sigma_y}\rangle$ respectively. 

\begin{figure}[ht!]
\centering
\includegraphics[scale=0.5]{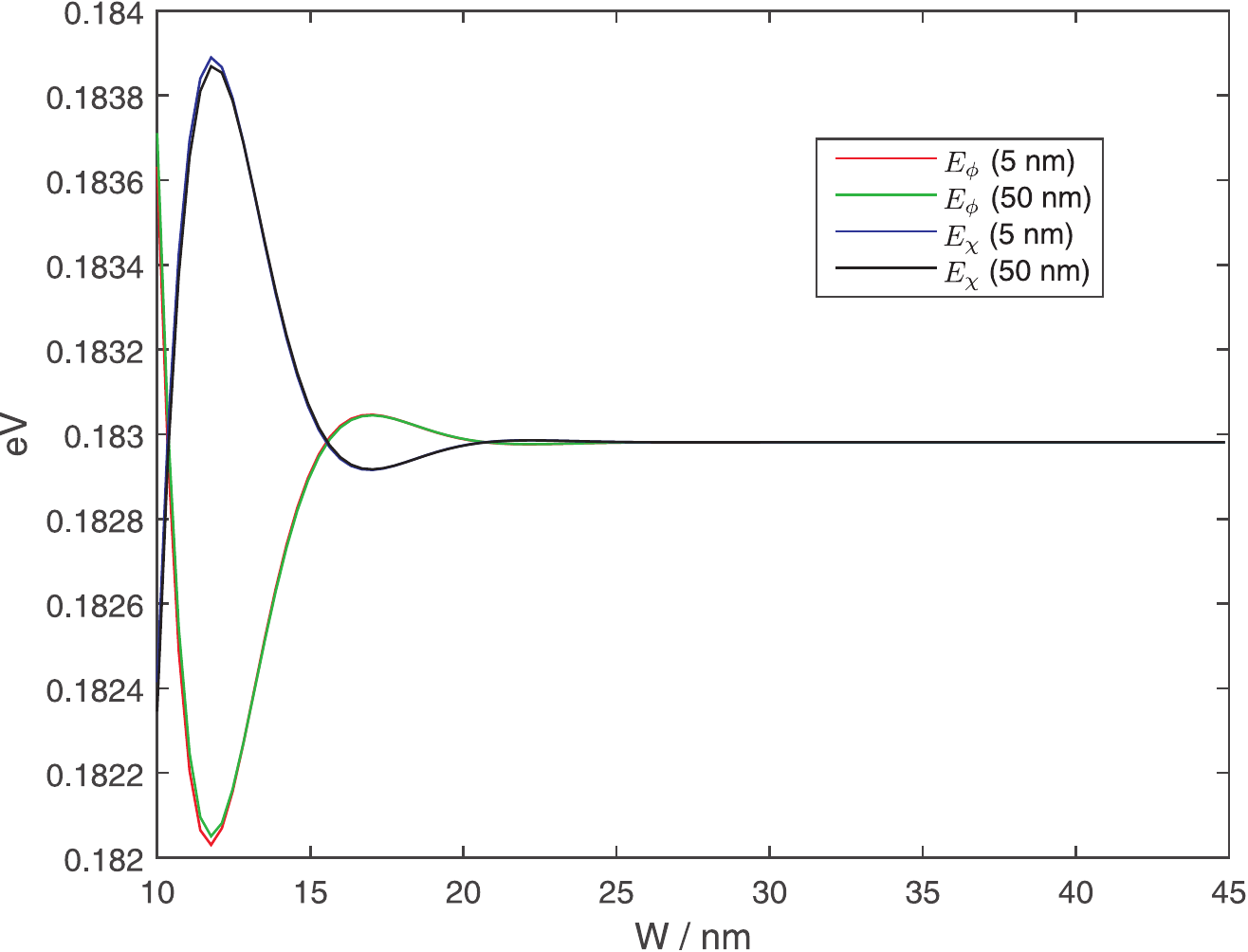}
\caption{ $E_{\phi}$ and $E_{\chi}$ as a function of the nanotube width $W$ for two different values of $R_i = 5\ \mathrm{eV}$ and $50\ \mathrm{eV}$ as indicated in the legend.   } 
\label{gEphichi}
\end{figure}

The eigenenergies of $|\varphi_{\pm \sigma_\phi} \rangle$ and  $|\chi_{\pm \sigma_\phi}\rangle$ states, which we  also label as $E_\phi$ and $E_\chi$ respectively, are shown in Fig. \ref{gEphichi} for the smallest and largest values of $R_i$ considered here.  The energies are, to a good approximation, independent of $R_i$ and equal to the corresponding eigenenergies for the perpendicular Hamiltonian eigenstates of the \textit{flat} thin film.  

The close resemblance between the eigenstates of the flat and curved perpendicular Hamiltonian is perhaps unsurprising. The neighborhood of a point on the surface of a cylinder tends to that of a point on a flat surface in the limit $r \rightarrow \infty$.  The combination $\sqrt{r}\psi$ appears in the calculation of expectation values in cylindrical coordinates. In  calculating the integral in the expectation value $\langle \Psi|O|\Phi \rangle = \int \mathrm{d}r\ r\Psi^*O\Phi $, the factor of $r$ can be split between the $|\Psi\rangle$ and $|\Phi\rangle$ wavefunctions as $= \int \mathrm{d}r\ (\sqrt{r}\Psi^*)O(\sqrt{r}\Phi)$.   This resembles the corresponding integral in a flat surface $\int \mathrm{d}y \Psi'^* O \Phi'$ with the identification of $y \rightarrow r$, $\Psi' \rightarrow \sqrt{r}\Psi$ and  $\Phi' \rightarrow \sqrt{r}\Phi$.  

\section{Parallel Hamiltonian} 
The parallel Hamiltonian for the TI nanotube $H_{(4B), cy, \parallel}$  in the lab frame reads 
\begin{eqnarray*}
	H_{(4B), cy, \parallel} &=& \left( C_1 k_z^2 + C_2 \frac{k_\phi^2}{r^2} \right)\mathbb{I}_4 +  \\ 
	&& \left( M_1 k_z^2 + M_2 \frac{k_\phi^2}{r^2} \right)t_z\mathbb{I}_\sigma \\
	&& + B k_z t_y\mathbb{I}_\sigma + A \frac{k_\phi}{r} t_x\sigma_r.
\end{eqnarray*}

In order to derive an effective Hamiltonian for the surface states, we now now take the projection of $H_{(4B),cy,\parallel}$ with respect to the four basis states  $|\varphi_{\pm \sigma_\phi}\rangle$ and $|\chi_{\pm \sigma_\phi} \rangle$.  The eigenstates of $H_{(4B),cy,\perp}$ calculated numerically in the previous section are in the rotated frame. We thus need to perform a unitary transformation on $H_{(4B), cy, \parallel}$ in order to take its projection with the numerically calculated  $H_{(4B),cy,\perp}$ eigenstates. The resulting effective Hamiltonian is in the rotated frame.  We then perform the inverse unitary transformation to convert the effective Hamiltonian back to the lab frame.  

In the course of calculating the projections of $H_{(4B), cy, \parallel}$ on the basis states, we will be integrating out the factors of $\frac{1}{r}$ that occur in $\frac{1}{r^2}\partial_\phi^2$ in the Laplacian operator, as well as in $\frac{1}{r}\partial_\phi$.  Counting the factor of $\sqrt{g}=r$ in the infinitesimal cross section area element $r \mathrm{d}r$ as well,  the integrands resulting from terms not containing $k_\phi$ and $k_\phi^2$ will contain a factor of $r$, the $k_\phi$ terms will contain no factors of $r$ while those from $\nabla^2$ will contain a factor of $\frac{1}{r}$. (In contrast, for a \textit{flat} thin film with normal in the  $x$ direction, the $x$ coordinate does not appear explicitly as a multiplicative factor in any of the integrals.)   The integrands containing a factor of $r$ resemble the integrands occurring for a flat film where the integrands have even or odd parity. The integrals with odd parity evaluate to 0. The integrals containing other powers of $r$,  do not obey these simple symmetry relations and do not cancel out exactly. Compared to the flat TI film, these additional terms give rise to more non-zero terms in the effective Hamiltonian for the cylindrical thin film. 

Besides the real spin degree of freedom represented by the $\pm \sigma_\phi$ states, the two `types' of eigenstates,  $|\varphi\rangle$ and $|\chi\rangle$, lead to an additional two-state degree of freedom which we denote as $\tau$ with $\tau_z \equiv |\chi\rangle\langle\chi|-|\varphi\rangle\langle\varphi|$ and analogously for $\tau_x$ and $\tau_y$. The $+\tau_x,\pm \sigma_\phi$ polarized state is thus  $\frac{1}{\sqrt{2}} (|\varphi_{\pm \sigma_\phi}\rangle + |\chi_{\pm \sigma_\phi}\rangle ). $ In particular, a visual inspection (not shown)  of the $\tau_i$ polarized wavefunctions indicates that  $\tau_x$ polarization carries the physical significance of indicating whether the eigenstates are localized nearer the inner ($+\tau_x$) or outer ($-\tau_x$) radius.

\subsection{Terms resulting from $\phi$ derivatives} 
In rotating $H_{(4B), cy, \parallel}$ to the lab frame,  the terms containing $k_\phi$ do not commute with $U$. We mentioned in the discussion following Eq. \ref{UkrU} that a part of the commutator between $k_\phi$ and $U$ went into contributing the $\frac{1}{r}$ factor inside the $-i(\partial_r + \frac{1}{r})$ terms in the perpendicular Hamiltonian, and that the remainder of the commutator is a $\frac{A}{2r} t_x \tilde{\sigma}_y$ term. The latter has not been included in our $H_{(4B), cy \perp}$ and will be considered here.  Putting this and the terms containing $k_\phi$ together and projecting to our four basis states, we have, in the rotated frame, the terms 
\begin{eqnarray*}
	&& \sum_{\substack{\Psi,\Psi' = \varphi,\chi \\ s,s'=\pm}} |\Psi, s\rangle \langle \Psi, s | A \big( U  \frac{k_\phi}{r} \tau_x\sigma_r   )U^\dagger + \\ 
	&&\  \frac{1}{2r} t_x \tilde{\sigma_y} \big)  |\Psi',s'\rangle\langle \Psi',s| \\ 
	&\simeq& -\frac{A}{2} ( \mathrm{Re}(M'^{(0)}_x) \tilde{\sigma}_x \tau_y + \mathrm{Im} (M'^{(0)}_x) \tilde{\sigma}_x \tau_x ) \\
	&-&i \frac{A}{2} \Big(  ( C'^{(0)}_x + P'^{(0)}_x)\tilde{\sigma}_x\mathbb{I}_\tau + ( C'^{(0)}_x - P'^{(0)}_x ) \tilde{\sigma}_x\tau_z + \\
	&& 2 M'^{(0)}_x \tilde{\sigma_x}\tau_x \Big) k_\phi.
\end{eqnarray*}
where $C^{(n)}_i \equiv \int \mathrm{d}r\ r^n \chi_{+\sigma_\phi}^\dagger t_i\chi_{+\sigma_\phi}$, $P^{(n)}_i \equiv \int \mathrm{d}r\ r^n \varphi_{+\sigma_\phi}^\dagger t_i\varphi_{+\sigma_\phi}$ and $M^{(n)}_i \equiv \int \mathrm{d}r\ r^n \chi_{+\sigma_\phi}^\dagger t_i\varphi_{+\sigma_\phi}$ (`C',`P' and `M' for \textit{c}hi, \textit{p}hi and \textit{m}ixed respectively). We have also defined $C'^{(n)}_i \equiv \int \mathrm{d}r\ r^n \chi_{+\sigma_\phi}^\dagger\sigma_i\chi_{-\sigma_\phi}$ and the primed versions of $P$ and $M$ are defined analogously where the integrand contains a $+\sigma_\phi$ bra and a $-\sigma_\phi$ ket. In deriving the above, we made use of the fact that $\langle \Phi,-\sigma_\phi|t_x|\Psi,+\sigma_\phi \rangle = - \langle \Phi,+\sigma_\phi|t_x|\Psi,-\sigma_\phi \rangle$ where $\Phi$ and $\Psi$ can each be either one of $\varphi$ and $\chi$ and $i = x,y$. We shall, in deriving the expressions encountered later, also make use of the identities $\langle \Phi,+\sigma_\phi|\sigma_j|\Phi,+\sigma_\psi \rangle =  \langle \Phi,-\sigma_\phi|\sigma_j|\Phi,-\sigma_\phi \rangle$ for $j=\mathbb{I},z$.

The terms containing $\nabla^2$ also do not commute with $U$ due to the presence of the $\frac{1}{r^2}\partial_\phi^2$ factor. The non-commutativity of $U$ and $\partial_\phi^2$ leads to the emergence of terms proportional to $\partial_\phi$ and constant terms. The latter terms \textit{do not} completely disappear after taking their projections with the four basis states and rotating back to the lab frame. The contributions of the parallel Hamiltonian terms containing $\nabla^2$ will be listed in our final expression for the lab frame effective surface state Hamiltonian Eq. \ref{Hpaokahliao}. 

\subsection{Terms resulting from $k_z$}

The portions of the effective Hamiltonian containing $k_z$ share the same form for both the cylindrical and flat thin films. We have, writing $C_z \equiv C^{(1)}_z$ for the cylindrical nanotube and $C_z \equiv \int \mathrm{d}x \chi^\dagger_{+y}t_z\chi_{+y}$ for the flat film, 

\begin{eqnarray*}
	&& \sum_{\substack{\Psi,\Psi' = \varphi,\chi \\ s,s'=\pm}} |\Psi, s\rangle \langle \Psi, s |  U   \big  (C_1\mathbb{I}_4 + M_1\mathbb{I}_\sigma t_z) k_z^2  + \\ 
	&&\    Bk_z t_y\big)U^\dagger |\Psi',s'\rangle\langle \Psi',s| \\ 
	&\simeq&  B k_z \Big( \mathrm{Re}( M_y) \tilde{\sigma}_z\tau_x - \mathrm{Im} (M_y)\tilde{\sigma}_z\tau_y)  \\
	&+&  (\frac{M_1}{2} ( C_z + P_z)) + C_1) \mathbb{I}_4 k_z^2 \\
	&+&  \frac{M_1}{2} ( C_z - P_z) \mathbb{I}_\sigma\tau_z k_z^2.
\end{eqnarray*} 

In writing the above, we used the approximation that $\sqrt{r}$ times the wavefunctions for the cylindrical system are almost identical to  the corresponding wavefunctions for the flat film. For the flat film, $\int \mathrm{d}x f_+f_- = 0$, so that terms which are proportional to it such as $\langle \chi_{\pm\sigma_y}|t_y|\chi_{\pm\sigma_y} \rangle$ evaluate to 0. The absence of such terms is one of the contributing factors to the relatively smaller number of terms containing $k_z$ compared to the terms containing $k_\phi$.
  
Since $k_z$ commutes with $U$, the unitary transformation of terms containing $k_z$ from the rotated frame back to the lab frame can be accomplished by changing the real spin operators $\tilde{\sigma}_z \rightarrow \sigma_\phi$ for the nanotube without introducing any additional terms. The corresponding terms for the flat TI thin film are obtained by changing $\tilde{\sigma}_z \rightarrow \sigma_\phi$.

The terms containing $k_y$ and $k_y^2$ in the \textit{flat} thin film have a similar form to those containing $k_z$ and $k_z^2$  -- 
\begin{eqnarray*}
	&& \sum_{\substack{\Psi,\Psi' = \varphi,\chi \\ s,s'=\pm}} |\Psi, s\rangle \langle \Psi, s |  U   \big  (C_2\mathbb{I}_4 + M_2\mathbb{I}_\sigma t_z) k_y^2  + \\ 
	&&\    A k_y t_y\big)U^\dagger |\Psi',s'\rangle\langle \Psi',s| \\
	&\simeq&A  \mathrm{Im}( M_x') \sigma_z\tau_x k_y \\
	&+&  \left(\frac{M_2}{2} ( C_z + P_z))+C_2 \right) \mathbb{I}_4 k_y^2 \\
	&+&  \frac{M_2}{2} ( C_z - P_z) \mathbb{I}_\sigma\tau_z k_y^2.
\end{eqnarray*}

Adopting the notation that $h_{\alpha\beta}$ are the terms independent of $k$ which go with $\tau_\alpha\sigma_\beta$, $v_{\alpha\beta\gamma}$ the terms which go with $k_\alpha\tau_\beta\sigma_\gamma$ and $\mu_{\alpha\beta\gamma}$ the terms which go with $k^2_\alpha\tau_\beta\sigma_\gamma$, the effective Hamiltonian for a flat thin film in the \textit{lab} frame (with a superscript of $(f)$ added to $h$, $v$ and $\mu$s to indicate that these are the parameters for a \textit{f}lat film) reads
\begin{eqnarray}
	H^{(f)} &=& h^{(f)}_{\mathbb{I}\mathbb{I}}\mathbb{I}_4 + h^{(f)}_{z\mathbb{I}}\tau_z\mathbb{I}_\sigma \nonumber \\ 
	&+&  v^{(f)}_{zxy}\tau_x \sigma_y  k_z + v^{(f)}_{yxz} \tau_x \sigma_z  k_y  \nonumber \\
	&+& \sum_{\alpha=(z,y)} (\mu^{(f)}_{\alpha \mathbb{I}\mathbb{I}} \mathbb{I}_4 + \mu^{(f)}_{\alpha z \mathbb{I}} \tau_z\mathbb{I}_\sigma) k_{\alpha}^2. \label{HeffFlat} 
\end{eqnarray}

The effective Hamiltonian for nanotube surface states is rather more complicated. Collecting all the terms and dropping those terms which are, to numerical precision 0 in our parameter range, the effective surface state Hamiltonian for the nanotube in the lab frame reads 
\begin{widetext}
\begin{eqnarray}
	&&H_{cy}  \nonumber \\ 
	&=& \frac{1}{4} ( 2C_2 F^{(-1)}_{\mathbb{I}+} + M_2F'^{(-1)}_{z+} - iA_0 F'^{(0)}_{x+} + 2(E_\varphi + E_\chi) )\mathbb{I}_4 \nonumber \\
	&& + \frac{1}{4} \big( -2i A_0 M'^{(0)}_x \tau_x + 2 M_2 M'^{(-1)}_z \tau_y + (2 C_2 F^{(-1)}_{\mathbb{I}-} + M_2 F'^{(-1)}_{z-}-i A_0 F'^{(0)}_{x-} + (E_\chi - E\varphi)) \tau_z \big) \mathbb{I}_\sigma  \nonumber \\
	&& + \frac{3}{4}C_2 \mathrm{Im}(M'^{(-1)}_x)\tau_x \sigma_r \nonumber \\
	&& + \frac{1}{4} \big( (-M_2 F'^{(-1)}_{z+} + i A_0 F'^{(0)}_{x+})\mathbb{I}_\tau - 2i A_0 M'^{(0)}_x \tau_x - 2M_2 M^{(-1)}_z \tau_y + (-M_2 F'^{(-1)}_{z-} + i A_0 F'^{(0)}_{x-})\tau_z \big) \sigma_z \nonumber \\
	&+& \frac{1}{2} \big( M_2 (F^{(-1)}_{z+} - F^{(-1)}_{\mathbb{I}+} )\mathbb{I}_\tau - M_2\mathrm{Re}(M^{(-1)}_{z})\tau_y + M_2 (F^{(-1)}_{z-} - F^{(-1)}_{\mathbb{I}-})\tau_z \big)\mathbb{I}_\sigma k_\phi + \nonumber \\
	&& -\frac{C_2}{4}\mathrm{Im}(M'^{(-1)}_{x}) \tau_x \{\sigma_r, k_\phi\} + \nonumber \\
	&& \frac{1}{2}\big( (C_2 F^{(-1)}_{\mathbb{I}+} + M_2 F'^{(-1)}_{z+} +  M_2 F^{(-1)}_{\mathbb{I}+} - i A_0 F'^{(0)}_{x+})\mathbb{I}_\tau  - 2i A_0 M'^{(0)}_x\tau_x \nonumber \\
	&& + 2M_2\mathrm{Re}(M^{(-1)}_z)\tau_y + (C_2 F^{(-1)}_{\mathbb{I}-} + M_2 F'^{(-1)}_{z-} +  M_2 F^{(-1)}_{\mathbb{I}-} - i A_0 F'^{(0)}_{x-})\tau_z\big)\sigma_z k_\phi \nonumber \\
	&+& \frac{1}{2} \big ( (C_2+M_2) F^{(-1)}_{\mathbb{I}+} \mathbb{I}_\tau -2 M_2 M^{(-1)}_z \tau_y +  (C_2+M_2) F^{(-1)}_{\mathbb{I}-}\tau_z \big )\mathbb{I}_s k_\phi^2 \nonumber \\
	&+& B_0 \mathrm{Re}{M^{(1)}_y}\tau_x \sigma_\phi k_z \nonumber \\
	&+& \frac{1}{2} ((M_1 F^{(1)}_{z+}+2C_1) \mathbb{I}_\tau  + M_1 F^{(1)}_{z-} \tau_z) k_z^2 \label{Hpaokahliao}
\end{eqnarray}
\end{widetext}
where we introduced the shorthand notation $F'^{(n)}_{i, \pm} \equiv C'^{(n)}_i \pm P'^{(n)}_i$. Note that we written the term containing $k_\phi$ and $\sigma_r$ in the symmeterized form $\{ v_\phi, \sigma_r\}$ because the two terms do not commute with each other.  Using a similar shorthand notation adapted in Eq. \ref{HeffFlat}, the effective Hamiltonian for the nanotube can be written as 
\begin{eqnarray}
	H =&&  h_{\mathbb{I}\mathbb{I}}\mathbb{I}_4 + \left(\sum_{\alpha=(x,y,z)} h_{\alpha\mathbb{I}}\tau_\alpha \mathbb{I}_\sigma\right)  \nonumber  \\ 
	&+&h_{xr}\tau_x\sigma_r + \left(\sum_{\alpha=(\mathbb{I},x,y,z)} h_{\alpha z}\tau_\alpha\sigma_z\right) \nonumber \\
	&+&\left(\sum_{\alpha=(\mathbb{I},y,z)} v_{\phi\alpha\mathbb{I}} \tau_\alpha\mathbb{I}_\sigma + \sum_{\alpha=(\mathbb{I},x,y,z)} v_{\phi\alpha z} \tau_\alpha\sigma_z\right) k_\phi  \nonumber \\
	&+& + v_{\phi xr} \tau_x \{ k_\phi,\sigma_r\} + v_{zx\phi} \tau_x\sigma_\phi k_z \nonumber \\
	&+&\sum_{\alpha=(z,\phi)} (\mu_{\alpha \mathbb{I}\mathbb{I}} \mathbb{I}_4 + \mu_{\alpha z \mathbb{I}} \tau_z\mathbb{I}_\sigma) k_{\alpha}^2. \label{HeffCy} 
\end{eqnarray}

Some of the terms in Eq. \ref{HeffCy} for the nanotube have direct analogs in Eq. \ref{HeffFlat} for the flat film. The terms containing the $\mu$s, $h_{\mathbb{I}\mathbb{I}}$ and $h_{z\mathbb{I}}$ are direct analogs, while $v^{(f)}_{zxy} k_z\tau_x\sigma_y \leftrightarrow v_{zx\phi} k_z\tau_x\sigma_\phi$ and $v^{(f)}_{yxz}k_y\tau_x\sigma_z  \leftrightarrow v_{\phi x z} k_\phi \tau_x\sigma_z$. The latter two terms give the usual Dirac fermion Hamiltonian $v (\vec{p}\times\hat{n})\cdot\vec{\sigma}$ for TI surface states and reflect the well known fact that $v$ has opposite signs for the two surfaces.  

The terms which do not have analogs in between the flat thin film and nanotube, or which have additional contributions in the nanotube, can be attributed to a combination of the position dependence of the surface normal $\hat{n}$ (which in affects the spin orbit interaction field) and the asymmetry between the inner and outer surfaces of the nanotube. For example, the powers of $r$ indicated by the superscript bracketed index $n$ in $F^{(n)}$, $C^{(n)}$ etc. in Eq. \ref{Hpaokahliao} in the $h$ terms give an indication of where these terms come from. The terms with $n=-1$ originate from the non-commutativity of $k_\phi^2/r^2$ in the Laplacian operator in $H_{\parallel, cy}$ with $U$, and those with $n=0$ from the non-commutativity of $k_\phi/r$ with $U$. The non commutativity is reflective of the position dependence of the surface normal, which leads to the direction of the spin orbit interaction field varying around the circumference of the tube. Further, unlike the flat TI case where  $\int \mathrm{d}z f_+f_- = 0$ results in integrals like $C_{x/y}$ and $P_{x/y}$ evaluating to 0, the asymmetry between the inner and outer radius in turn results in integrals that arise from the non-commutativity with $U$ like  $C^{(n)}_{x/y}$ and $P^{(n)}_{x/y}$ for $n \neq 1$ evaluating to finite values. Similarly, the terms with $n=-1$ appearing in the $v_{\phi}$ terms in Eq. \ref{Hpaokahliao} also originate from the non-commutativity of $k_\phi^2/r^2$ with $U$.

\section{Results} 

\begin{figure}[ht!]
\centering
\includegraphics[scale=0.3]{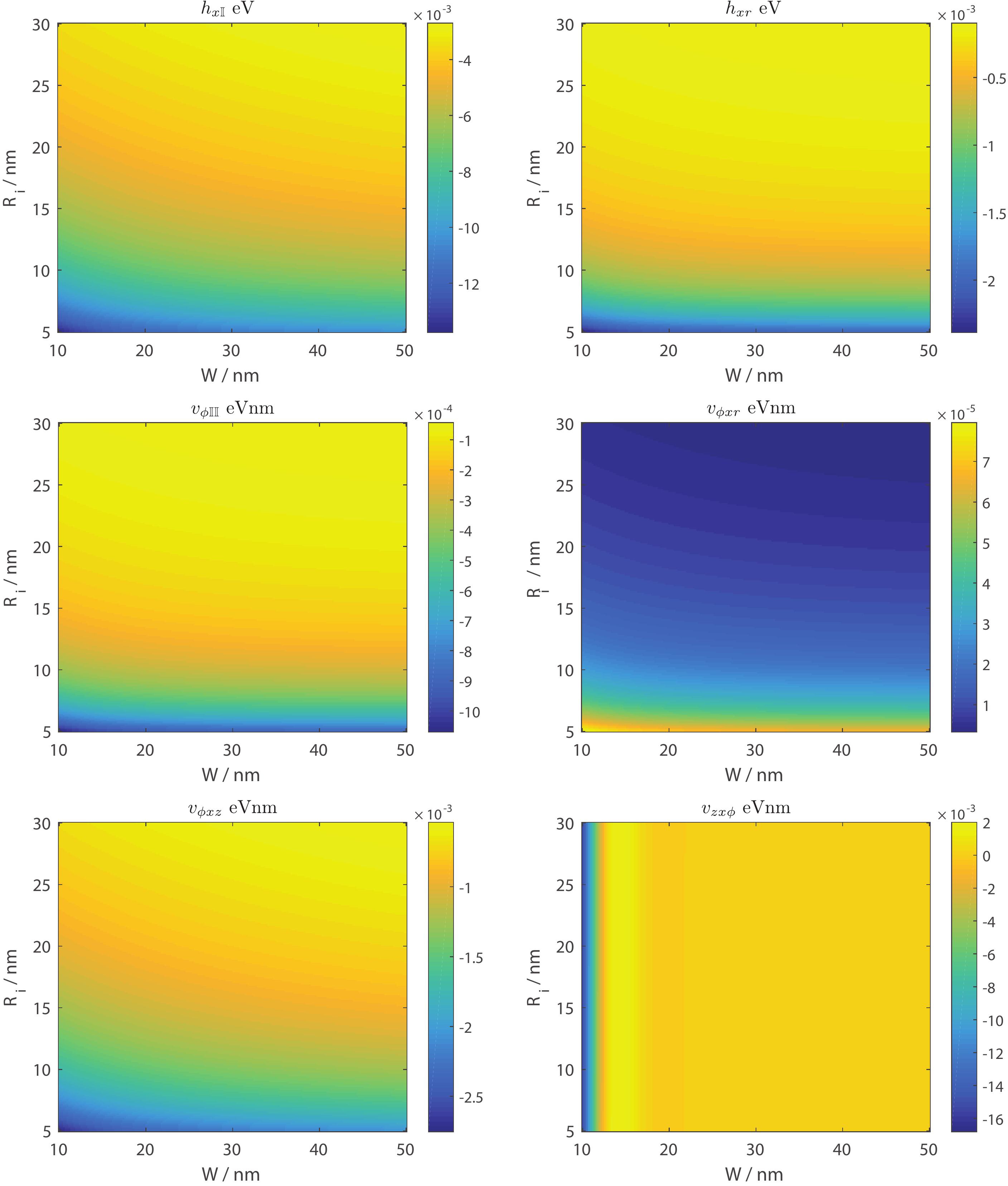}
\caption{   The numerical values for some of the parameters in Eq. \ref{HeffCy} as a function of the nanotube inner radius and width.}
\label{gParams}
\end{figure}

Fig. \ref{gParams} shows the values of some of these parameters for various values of inner radius and nanotube widths. The parameters shown here have the largest magnitudes for the $h$s and $\mu$s which go with each direction of $\sigma$.  $h_{xz}$ is, to numerical precision, equal to $h_{x\mathbb{I}}$ despite the differing forms of the underlying expressions. $h_{\mathbb{I}\mathbb{I}}$ (not shown here) also has a rather large magnitude of around $0.185\ \mathrm{eV}$ for the parameter ranges shown here but is relatively unimportant as it amounts to a constant energy shift. The $\mu$s are not shown in the figure as the plots of their magnitudes are similar to the quantities already shown.  Amongst the $\mu_\phi$s, $\mu_{\phi\mathbb{I}\mathbb{I}}$ is at least 3 times larger in magnitude than the next largest $\mu_\phi$ ($\mu_{\phi y\mathbb{I}}$). Its plot is similar to that of $v_{\phi x z}$ except that the scale bar goes from slightly more than 0 to $11\times 10^{-3}\mathrm{eV}$. Amongst the $\mu_z$s,  $\mu_{z\mathbb{I}\mathbb{I}}$ has the largest magnitude of at least 10 times bigger than the next largest $\mu_z$. The plot is similar to that of $v_{zx\phi}$ with the scale bar taking values from $-4.68 \mathrm{eV}$ to $-4.69 \mathrm{eV}$. 

The variation of the Hamiltonian parameters with $W$ and $R_i$ fall into two broad categories. 

$v_{zx\phi}$ in the figure exemplifies the first of these two categories where there is a very weak dependence on $r_i$, an oscillatory dependence on $W$ for $W$ less than around $25 \mathrm{eV}$ and a constant value for larger values of $W$. This behavior is also exhibited by $h_{z\mathbb{I}}$, $\mu_{z\mathbb{I}\mathbb{I}}$ and $\mu_{zx\mathbb{I}}$. The variation of $\mu_{\phi z \mathbb{I}}$ and $v_{\phi z z}$ also fall into this category but have a stronger dependence on $R_i$. The oscillatory variation of these parameters with $W$ at small $W$ may be related to the variation of $E_{\chi}$ and $E_{\varphi}$ with $W$, as shown in Fig. \ref{gEphichi}.  

The variation of the remaining parameters fall into the second category where there is a stronger dependence on $r_i$ than on $W$, and for which at large values of $W$ there is a relatively sharp jump in the parameter values at $R_i$ around $7\ \mathrm{nm}$. This dependence might be related to the asymmetry of the wavefunctions between the inner and outer surfaces of the nanotube which become especially evident at small values of $R_i$ relative to $W$. The asymmetry is further amplified when the wavefunctions are multiplied by inverse powers of $r$ in the evaluation of integrals such as $C^{(-1)}_i$. 

\begin{figure}[ht!]
\centering
\includegraphics[scale=0.4]{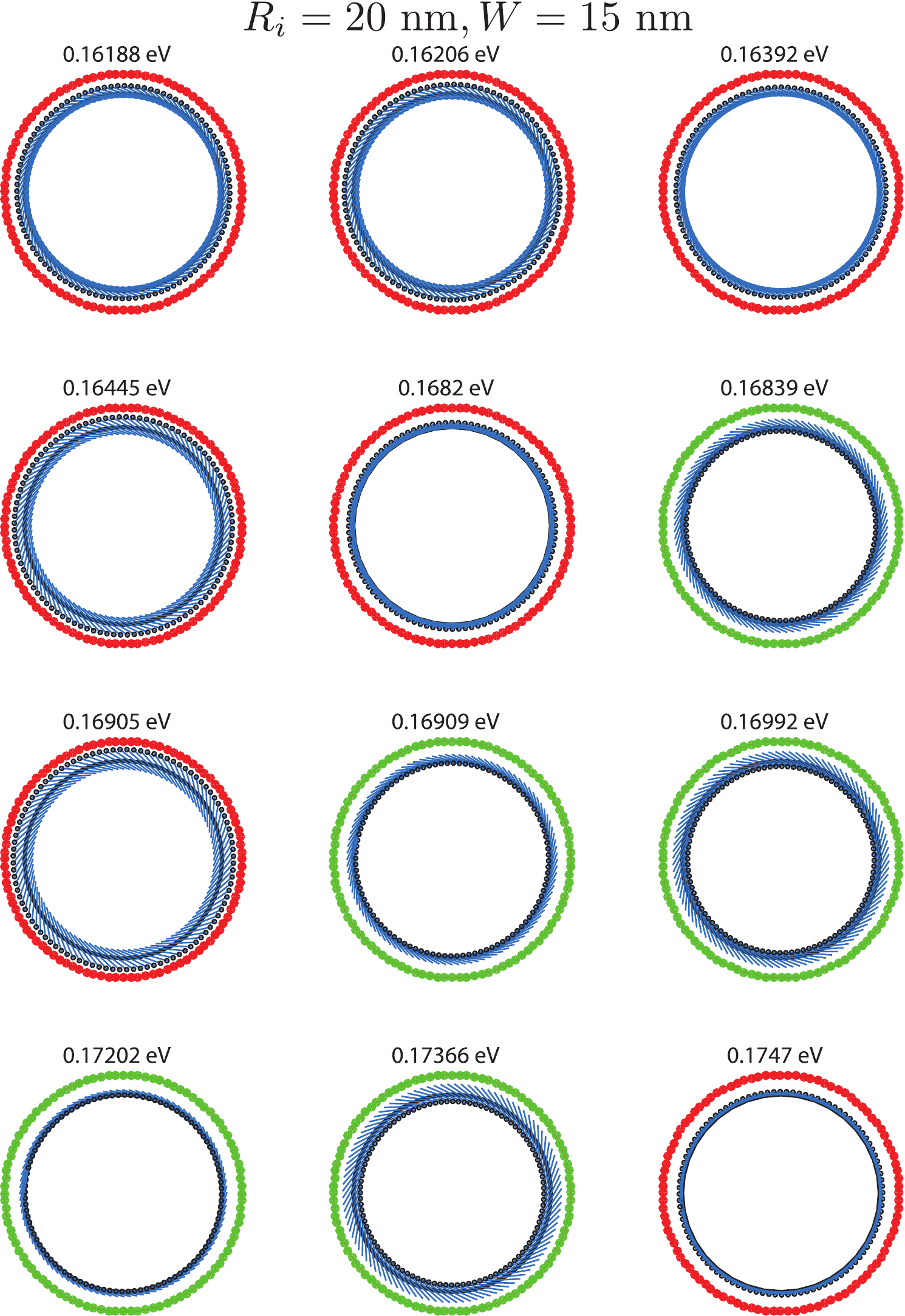}
\caption{  The twelve lowest energy eigenstates for a nanotube of width $15 \mathrm{nm}$ and inner radius $20 \mathrm{nm}$ at $k_z = 0.01 \mathrm{nm}^{-1}$. The direction of the real spin polarization on the $(xy)$ plane at each point along the circumference of the tube are denoted by the arrows scattered along the circumference with the lengths of the arrows being proportional to the magnitude of the in-plane polarization. The red / green dots denote the sign of $\langle \tau_x \rangle$ with red (green) dots denoting positive (negative) values of $\langle \tau_x \rangle$ which in correspond to states localized along the inner (outer) walls of the cylinder.  }        
\label{gR200W150}
\end{figure}

\begin{figure}[ht!]
\centering
\includegraphics[scale=0.4]{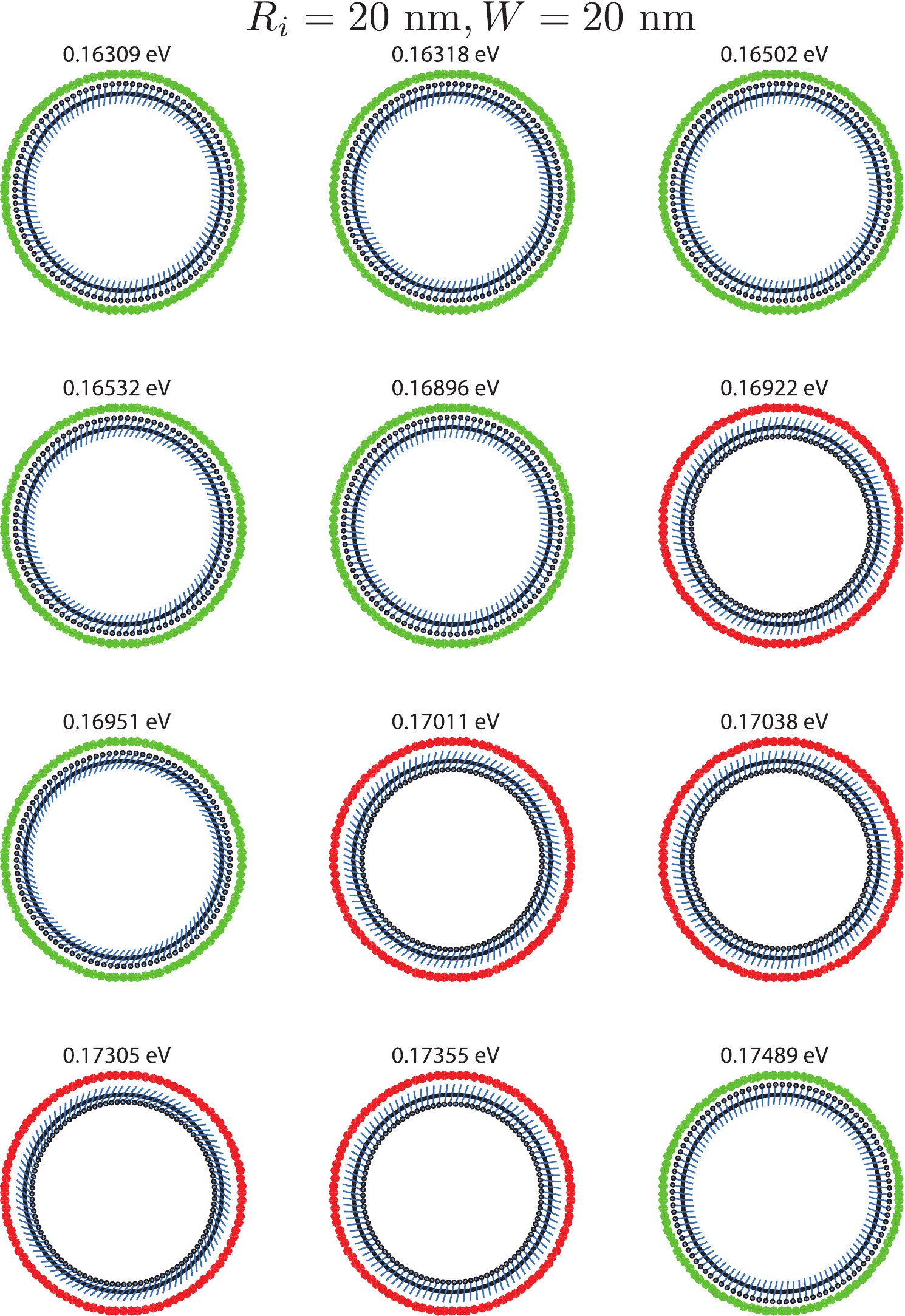}
\caption{  The twelve lowest energy eigenstates for a nanotube of width $20 \mathrm{nm}$ and inner radius $20 \mathrm{nm}$ at $k_z = 0.01 \mathrm{nm}^{-1}$.  }       
\label{gR200W200}
\end{figure}

\begin{figure}[ht!]
\centering
\includegraphics[scale=0.4]{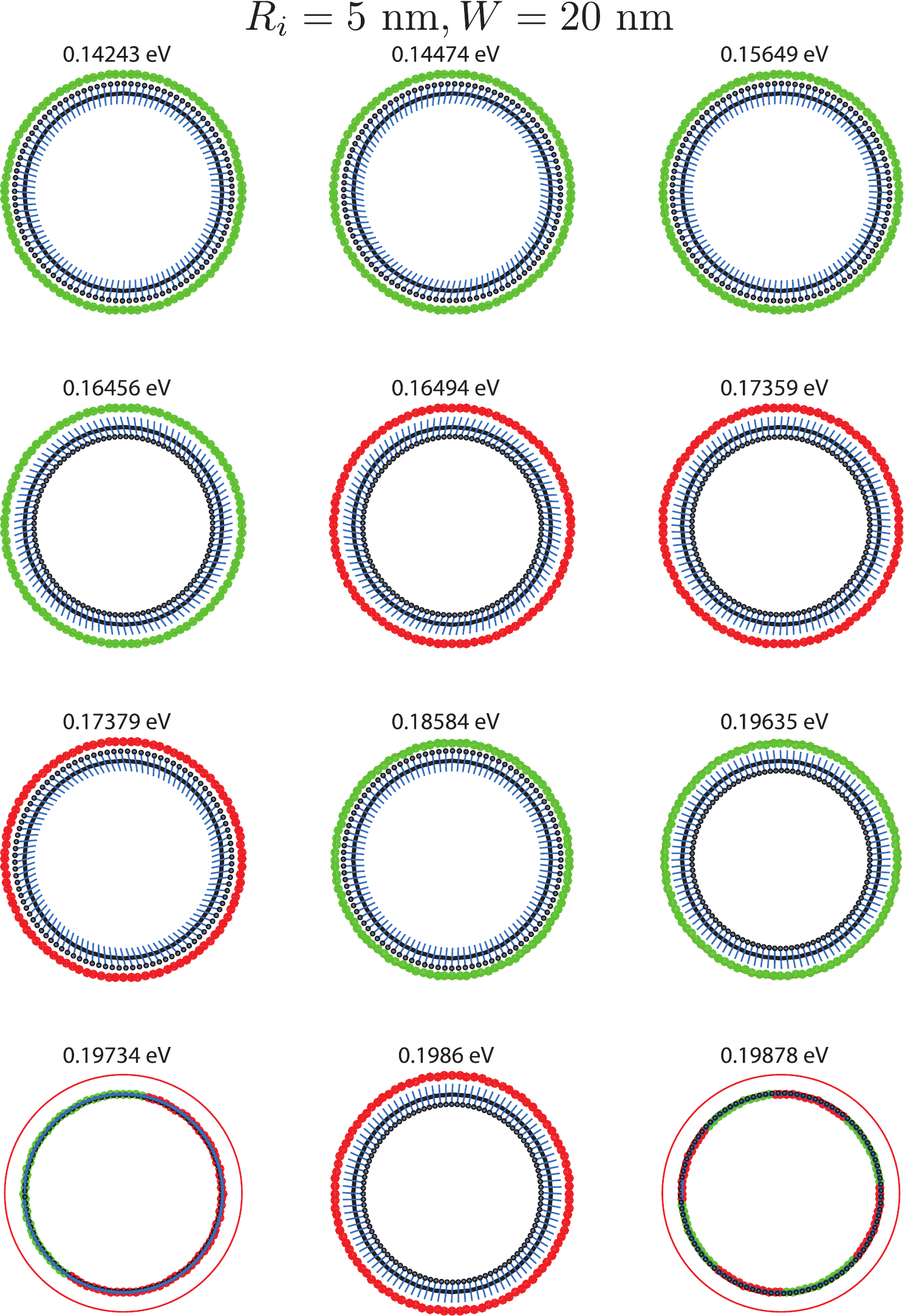}
\caption{  The twelve lowest energy eigenstates for a nanotube of width $20 \mathrm{nm}$ and inner radius $5 \mathrm{nm}$ at $k_z = 0.01 \mathrm{nm}^{-1}$.  }        
\label{gR50W200}
\end{figure}

The competition between the various integrals present in some of the Hamiltonian parameters leads to a change in the signs of the parameters as $W$ and $r_i$ are varied. This in turn results in the reversal of correlations between the various degrees of freedom (momentum, $\tau$ and $\sigma$) in the low energy eigenstates of the effective Hamiltonian. We illustrate this by comparing the parameters and eigenstates of the effective Hamiltonians of two differing widths $15\ \mathrm{nm}$ and $20\ \mathrm{nm}$ and the same inner radius $5 \mathrm{nm}$, and the same width $20\ \mathrm{nm}$ and two differing inner radii $5\ \mathrm{nm}$ and $20\ \mathrm{nm}$. The table below shows the numerical values of the effective Hamiltonian parameters for these values of widths and inner radii. Figs. \ref{gR200W150},\ref{gR200W200} and \ref{gR50W200} show the corresponding real spin-$xy$ polarization, the $\tau_x$ polarization and the eigenenergies of the 12 lowest energy eigenstates evaluated at $k_z = 0.01 \mathrm{nm}^{-1}$.  

\begin{widetext}
\begin{tabular}{|l||c|c|c|}
\hline
Parameter & $R_i = 20 \mathrm{nm}$ & $R_i = 20 \mathrm{nm}$ & $R_i = 5 \mathrm{nm}$ \\
 & $W= 15 \mathrm{nm}$ & $W=20 \mathrm{nm}$ & $W=20 \mathrm{nm}$ \\
\hline
$h_{\mathbf{I} \mathbf{I} } / \mathrm{eV}$ & $0.1838$&$0.1837$&$0.1895$ \\ 
$h_{x \mathbf{I} } / \mathrm{eV}$ & $0.004627$&$-0.004387$&$-0.01173$ \\ 
$h_{y \mathbf{I} } / \mathrm{eV}$ & $9.179\times 10^{-5}$&$-1.087\times 10^{-4}$&$-0.00159$ \\ 
$h_{z \mathbf{I} } / \mathrm{eV}$ & $7.714\times 10^{-5}$&$-7.836\times 10^{-6}$&$-9.181\times 10^{-6}$ \\ 
$h_{x r } / \mathrm{eV}$ & $2.519\times 10^{-4}$&$-2.333\times 10^{-4}$&$-0.002052$ \\ 
$h_{\mathbf{I} z } / \mathrm{eV}$ & $-3.366\times 10^{-4}$&$-3.118\times 10^{-4}$&$-0.002742$ \\ 
$h_{x z } / \mathrm{eV}$ & $0.004627$&$-0.004387$&$-0.01173$ \\ 
$h_{y z } / \mathrm{eV}$ & $-9.179\times 10^{-5}$&$1.087\times 10^{-4}$&$0.00159$ \\ 
$h_{z z } / \mathrm{eV}$ & $-1.517\times 10^{-7}$&$2.59\times 10^{-8}$&$7.765\times 10^{-7}$ \\ 
$v_{\phi \mathbf{I} \mathbf{I} } / \mathrm{eV}$ & $-0.001133$&$-0.00105$&$-0.009231$ \\ 
$v_{\phi y \mathbf{I} } / \mathrm{eV}$ & $-1.836\times 10^{-4}$&$2.174\times 10^{-4}$&$0.00318$ \\ 
$v_{\phi z \mathbf{I} } / \mathrm{eV}$ & $3.325\times 10^{-6}$&$-2.425\times 10^{-7}$&$5.59\times 10^{-7}$ \\ 
$v_{\phi x r } / \mathrm{eV}$ & $-8.396\times 10^{-5}$&$7.778\times 10^{-5}$&$6.84\times 10^{-4}$ \\ 
$v_{\phi \mathbf{I} z } / \mathrm{eV}$ & $0.001806$&$0.001673$&$0.01472$ \\ 
$v_{\phi x z } / \mathrm{eV}$ & $0.009254$&$-0.008774$&$-0.02346$ \\ 
$v_{\phi y z } / \mathrm{eV}$ & $1.836\times 10^{-4}$&$-2.174\times 10^{-4}$&$-0.00318$ \\ 
$v_{\phi z z } / \mathrm{eV}$ & $8.143\times 10^{-7}$&$-1.39\times 10^{-7}$&$-4.167\times 10^{-6}$ \\ 
$\mu_{\phi \mathbf{I} \mathbf{I} } / \mathrm{eV}$ & $0.001133$&$0.00105$&$0.009231$ \\ 
$\mu_{\phi y \mathbf{I} } / \mathrm{eV}$ & $1.836\times 10^{-4}$&$-2.174\times 10^{-4}$&$-0.00318$ \\ 
$\mu_{\phi z \mathbf{I} } / \mathrm{eV}$ & $5.108\times 10^{-7}$&$-8.72\times 10^{-8}$&$-2.614\times 10^{-6}$ \\ 
$v_{z x \phi } / \mathrm{eVnm}$ & $-0.001875$&$-1.841\times 10^{-4}$&$-1.859\times 10^{-4}$ \\ 
$\mu_{z \mathbf{I} \mathbf{I} } / \mathrm{eVnm}^2$ & $0.01054$&$0.01054$&$0.01054$ \\ 
$\mu_{z z \mathbf{I} } / \mathrm{eVnm}^2$ & $4.157\times 10^{-4}$&$-4.08\times 10^{-5}$&$-4.12\times 10^{-5}$ \\  
\hline
\end{tabular}
\end{widetext}

These three choices of nanotube dimensions illustrate the  differing behavior of the low energy eigenstates of nanotubes with the changes in the relative signs of the various parameters in the Hamiltonian as the inner and outer radii are varied. We first draw attention to the fact that $v_{zx\phi}$ has the same sign for all three nanotubes. The tubes plotted all have the same positive value of $k_z$, and a positive (negative) sign of $\langle\tau_x\rangle$ is always associated with a positive (negative) $\langle\sigma_\phi\rangle$. The $15\ \mathrm{nm}$ wide tube has opposite signs of $h_{x\mathbb{I}}$ relative to the two wider tubes. This results in the first few lowest energy eigenstates (where the contributions of $k_\phi$ is minimal) of the $15 \mathrm{nm}$ tube having an opposite sign of $\langle \tau_x \rangle$ relative to the other tubes.  The $15 \mathrm{nm}$ wide tube also has an opposite sign of $v_{\phi xr}$ from the other tubes. Thus whereas a positive (negative) $\langle \sigma_r \rangle$ occurs together with a positive (negative) $\langle \tau_x \rangle$ in this tube, the converse is true for all the eigenstates shown for the $W=20\ \mathrm{nm}, R_i=20\ \mathrm{nm}$ nanotube in Fig. \ref{gR200W200}, and most of the eigenstates of the  $W=20\ \mathrm{nm}, R_i=5\ \mathrm{nm}$ tube shown in Fig. \ref{gR50W200}.

The   $W=20\ \mathrm{nm}, R_i=5\ \mathrm{nm}$ tube displays an interesting phenomenon absent in the wider tubes -- the in-plane real spin and $\tau_x$ polarizations are almost 0 for two of the eigenstates. One possible reason for this might be due to the fact that in the other two tubes the magnitude of $v_{zx\phi}$ is larger than that of $v_{\phi x r}$ whereas in this tube the converse is true so that the competition between the energy contributions of these two terms may result in it being more energetically favorable to have almost 0 $\sigma_r$ and $\tau_x$ polarization.

\begin{figure}[ht!]
\centering
\includegraphics[scale=0.4]{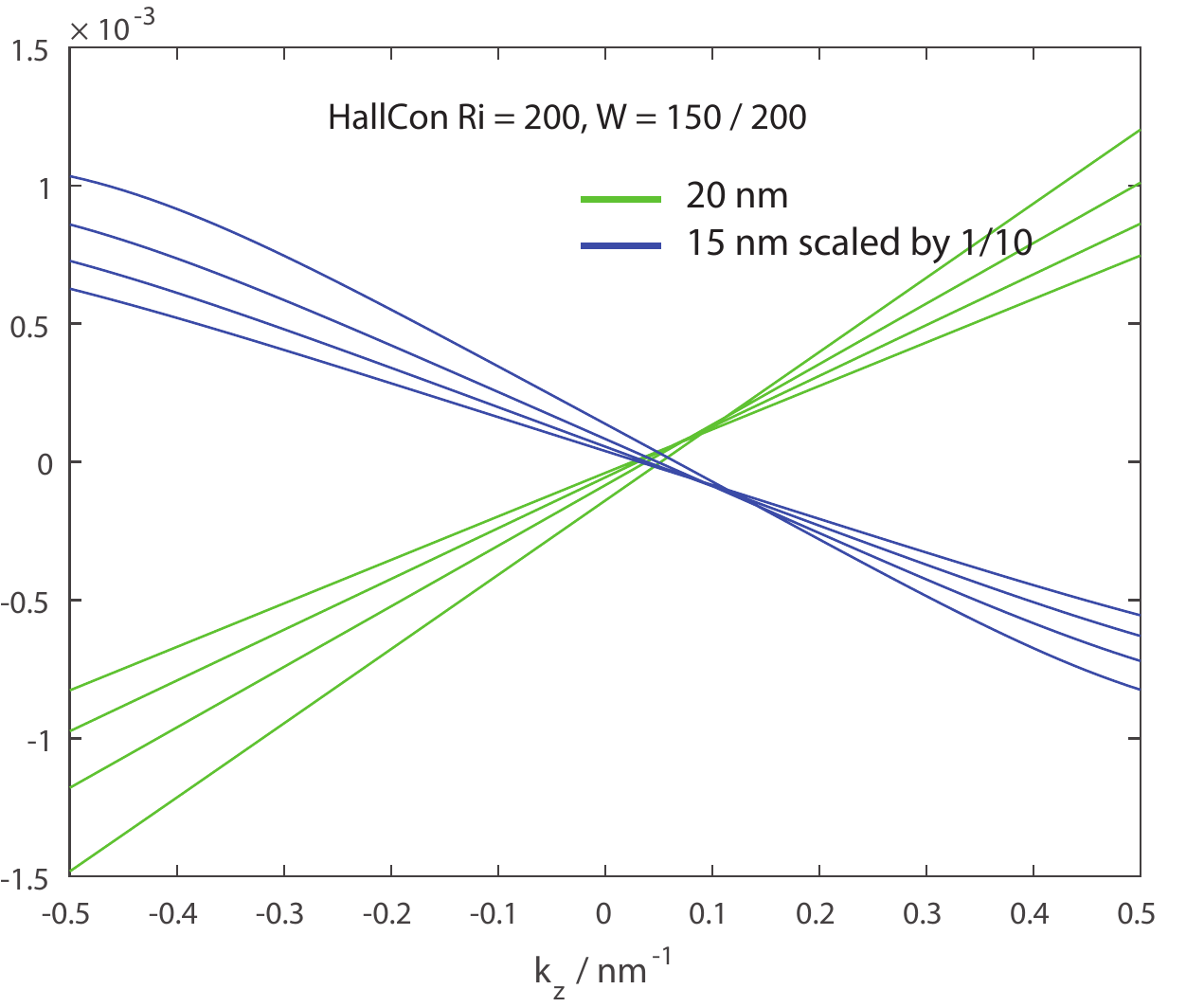}
\caption{  $\sigma_{\phi, z}$ for the four lowest energy states of the $15\ \mathrm{nm}$ and $20\ \mathrm{nm}$ wide tubes of inner radius $20\ \mathrm{nm}$ as a function of $k_z$. The values for the $15\ \mathrm{nm}$ tube has been scaled down by 1/10 in order to fit into the same vertical axis values.     }        
\label{gHallCon}
\end{figure}

The opposing sign of $v_{\phi xr}$ between the $15\ \mathrm{nm}$ and $20\ \mathrm{nm}$ wide tubes is also reflected in the Hall conductivity $\sigma_{\phi, z}$ relating the current flowing around the azimuthal $\phi$ direction due to an electric field in the $z$ direction calculated using the standard Kubo formula. Fig. \ref{gHallCon} shows that the conductivity for the four lowest energy states of the two widths have the opposite dependence on $k_z$ -- the $15\ \mathrm{nm}$ ($20\ \mathrm{nm}$ one increases (decreases) with $k_z$.  

\section{Conclusion} 
In this work we derived the effective Hamiltonian for the surface states of a TI nanotube with both an inner and outer surface. We showed that the combination of the position dependence of the surface normal around the circumference of the tube and the asymmetry between the inner and outer radius of the tube give rise to more terms in the TI nanotube absent in a flat thin film. The variation of the relative signs and magnitudes of the various parameters in the Hamiltonian as the inner radius and tube width give rise to differing behavior in the nanotube surface states.

\end{document}